\documentclass[11pt,a4paper]{article}
\usepackage{jcappub}
\usepackage{color}
\usepackage[T1]{fontenc}
\usepackage{ifthen}
\usepackage{booktabs}

\newcommand{\beqra}{\begin{eqnarray}}
\newcommand{\eeqra}{\end{eqnarray}}
\newcommand{\beq}{\begin{equation}}
\newcommand{\eeq}{\end{equation}}

\title{Improving Planck calibration by including frequency-dependent relativistic corrections}
\author[a]{Miguel Quartin}
\author[b,c]{and Alessio Notari}
\affiliation[a]{Instituto de F\'\i sica, Universidade Federal do Rio de Janeiro, 21941-972, Rio de Janeiro, RJ, Brazil}
\affiliation[b]{Departament de F\'isica Fondamental i Institut de Ci\'encies del Cosmos,
Universitat de Barcelona, Mart\'i i Franqu\'es 1, 08028 Barcelona, Spain}
\affiliation[c] {Dip. di Fisica, Universit\`a di Ferrara and INFN Sez. di Ferrara, Via Saragat 1, I-44100 Ferrara, Italy}

\abstract{The Planck satellite detectors are calibrated in the 2015 release using the ``orbital dipole'', which is the time-dependent dipole generated by the Doppler effect due to the motion of the satellite around the Sun. Such an effect has also relativistic time-dependent corrections of relative magnitude $10^{-3}$, due to coupling with the ``solar dipole'' (the motion of the Sun compared to the CMB rest frame), which are included in the data calibration by the Planck collaboration. We point out that such corrections are subject to a frequency-dependent multiplicative factor. This factor differs from unity especially at the highest frequencies, relevant for the HFI instrument. Since currently Planck calibration errors are dominated by systematics, to the point that polarization data is currently unreliable at large scales, such a correction can in principle be highly relevant for future data releases. }

\newpage

\begin{document}

%\flushbottom

\maketitle

\section{Summary of CMB Calibration}

The Planck satellite~\cite{Ade:2013sjv,Adam:2015rua} has measured CMB intensity maps at several different frequencies with both the Low Frequency Instrument (LFI) and the High Frequency Instrument (HFI). The detectors are calibrated using a given known source, which allows to fix the proportionality constant (the {\it gain} factor) between the detector's response and a known intensity value. In the 2013 release~\cite{Ade:2013sjv,Aghanim:2013bta,Ade:2013eta} such a source was the dipolar temperature pattern induced by the Doppler boosting of the primordial temperature monopole due to the velocity  $\boldsymbol{\beta_S}$ of the Sun with respect to the CMB rest frame, except for the two highest frequency channels which were calibrated on planets.

Such a dipole, called the ``solar dipole'' by the Planck collaboration, is practically time-independent during the observation time. Its amplitude $\beta_S$ is of order $10^{-3}$ in units of the speed of light, \emph{assuming} that the measured CMB dipole is mostly due to the Sun velocity compared to the CMB. This is generally considered a safe assumption and can be directly tested through measurements of the aberration of the CMB~\cite{Kosowsky:2010jm,Amendola:2010ty,Notari:2011sb}; such a test was performed on the Planck 2013 data~\cite{Aghanim:2013suk}, with still quite large error bars $\sim40\%$. In any case its value is not known independently {\it a priori} in order to calibrate the instrument (which poses a big disadvantage to the method) and for this reason the Planck team had to use the value measured by WMAP5~\cite{Hinshaw:2008kr,2011ApJS..192...14J}, which is itself subject to WMAP calibration uncertainties.

In the 2015 release~\cite{Adam:2015rua,Adam:2015vua,Ade:2015ads} such a source was replaced by the Doppler effect due to the motion with velocity $\boldsymbol{\beta_O}$ of the Planck satellite around the Sun, called the ``orbital dipole''. Although it is one order of magnitude smaller than the solar dipole (${\beta_O} = 1.0 \times 10^{-4}$) it has the great advantages of both being time-dependent (on a scale of 1 year) and having a very well known amplitude and direction. The very small uncertainties come only from the motion of the satellite inside the solar system, which is known to very high accuracy. The uncertainties are of the order $10^{-10}c$, which corresponds to just one part in a million. For this reason the 2015 release is thought to be more accurate and  the new absolute calibration of the Planck 2015 HFI instrument is higher by $2\%$ (in power) compared to 2013,  resolving the calibration differences  noted between WMAP and Planck, which goes down from $2.6\%$ in 2013 to $0.3\%$ in 2015.

These improvements led to shifts in the cosmological parameter likelihood. Although there were other improvements in Planck 2015 compared to 2013, according to the Planck collaboration calibration was the most important factor regarding the combination of the amplitude of the power spectrum and reionization optical depth: $A_s e^{-2\tau}$~\cite{Adam:2015rua}. In fact, its value was corrected upwards by around $2\%$~\cite{Adam:2015rua} (see their Section 10), which corresponds to a large 3.5$\sigma$ shift to the best-fit value~\cite{Planck:2015xua} (see their Table I).

The extremely high accuracy of the orbital motion cited above means that the biggest uncertainty in the orbital temperature dipole comes from our knowledge of the CMB background temperature $T_0$ which is currently only known at the level of $0.02\%$~\cite{Fixsen:2009ug}. Since the orbital dipole is directly proportional to $T_0$, this imposes a limit of $0.02\%$ to the accuracy of this calibration technique. However the most accurately calibrated channel has an uncertainty of $0.07\%$~\cite{Adam:2015rua} so this is not currently a serious limitation. % but might be so in the future until better measurements of $T_0$ are made.

Since the orbital motion is known to such a good accuracy and in order to avoid systematic errors the Planck team has taken into account also of the subleading relativistic Doppler corrections, in addition to the leading order dipole term. In particular such effects couple the solar dipole to the orbital dipole and are of order $\beta_S \beta_O$ which represents therefore a \emph{time-dependent}  relative correction of about $10^{-3}$ compared to the leading dipole term, of order $\beta_O$. An effect of such size cannot a priori be neglected because the typical systematics of the calibration process are precisely of the same order, of about $0.1\%-1\%$, depending on the channel~\cite{Adam:2015rua,Adam:2015vua}.  Let us also note that WMAP was also calibrated using the orbital dipole~\cite{Hinshaw:2003fc,Hinshaw:2008kr}, and so any possible bias in this procedure could also have affected WMAP and  propagated also to the Planck 2013 release.

In this short note we point out that although the relativistic effects have been included in the 2015 Planck release~\cite{Adam:2015rua,Adam:2015vua,Ade:2015ads},\footnote{In 2013 they were not included as corrections to the solar dipole in HFI~\cite{Ade:2013eta}.  In LFI~\cite{Aghanim:2013bta} they were included [see eq.~(A.1) therein], though it was implemented with a wrong factor of 2 due to a bug in the code as reported in its appendix A.} they should be nevertheless multiplied by a frequency-dependent factor, in the same way as discussed in~\cite{Kamionkowski:2002nd,Chluba:2004cn,Sunyaev:2013coa,Notari:2015kla} (following the original results from~\cite{Sazonov:1999zp}) for the purpose of the subtraction of the Doppler quadrupole from the measured maps. This would imply an ${\cal O}(1)$ correction on the relativistic terms, which should be relevant as long as the relativistic terms were actually necessary in the calibration. The reason why such a factor arises is that the Planck detectors are measuring a signal proportional to intensity, and the intensity is not linearly related to the primordial temperature in the CMB rest frame. There is also a frequency-independent Dipole due to the non-relativistic Doppler effect plus \emph{frequency-dependent} relativistic corrections, as we are going to show.

\section{Corrections due to Linearizing Temperature}

Following closely~\cite{Notari:2015kla} we recall the effects of a boost with a velocity $\,\mathbf{v}/c=\boldsymbol{\beta}\,$. An observer in the CMB rest frame looking at the CMB black body signal along a direction $\boldsymbol{\hat{n}}$ would see a temperature $ T(\boldsymbol{\hat{n}})=T_0  + \varepsilon \, \delta T(\boldsymbol{\hat{n}}) $, where we assume $\varepsilon=10^{-5}$ and so  $ \delta T(\boldsymbol{\hat{n}})$ is of order 1. An observer in a boosted frame would instead see along a direction  $\boldsymbol{\hat{n}}'$ a temperature~\cite{Peebles:1968}:
\begin{equation}
    T'(\boldsymbol{\hat{n}}')=
    \frac{T(\boldsymbol{\hat{n}})}{\gamma(1-\boldsymbol{\beta} \cdot \boldsymbol{\hat{n}}')} \,. \label{Dopplerab}
\end{equation}
The multiplicative factor is the Doppler shift and the change in the apparent arrival direction of photons is the aberration effect~\cite{Burles:2006xf, Challinor:2002zh}. We split the velocity into a constant term due to the motion of the Sun with respect to the CMB and a time dependent one due to the orbit of the satellite around the Sun:
\begin{equation}
    \boldsymbol{\beta}(t) \equiv \boldsymbol{\beta_S} + \boldsymbol{\beta_O}(t).
\end{equation}
In what follows we will not write the time dependence explicitly aiming simplicity of notation.

A completely isotropic map is unaffected by aberration, which explains why its leading effect is at most ${\cal O}(\varepsilon \, \beta)$ ($\sim 10^{-8}$). The Doppler effect instead affects also an exactly isotropic map and in fact it correlates the monopole with higher multipoles, inducing a dipole of order $\beta$ and a $n^{\rm th}$-pole of order $\beta^n$, as it can be seen expanding the multiplicative factor in eq.~\eqref{Dopplerab}.

A detector in a boosted frame measures a signal at a given frequency $\nu'$ proportional to intensity:
\begin{equation}
    I'(\nu') = I (\nu) \left( \frac{\nu'}{\nu}\right)^3 = \frac{2 \nu'^3}{e^{\frac{h \nu }{k_B T(\boldsymbol{\hat{n}})}}-1}  \,.
\end{equation}
where $ \nu= \nu' \gamma(1-\boldsymbol{\beta} \cdot \boldsymbol{\hat{n}}') $. In the $\beta=0$ limit the two frames coincide and fluctuations in intensity are given at first order in $\varepsilon$ by:
\begin{equation}
    \delta I(\nu)\,\approx\, \frac{2 \nu ^4   e^{\frac{\nu }{\nu_0}}}{\nu_0^2 \left(e^{\frac{\nu }{\nu_0}}-1\right)^2}  \varepsilon  \, \delta T(\boldsymbol{\hat{n}})
    \,\equiv\, K  \varepsilon \, \frac{\delta T(\boldsymbol{\hat{n}})}{T_0} \,,
    \label{K}
\end{equation}
with $\nu_0 \equiv k_B T_0 / h = (56.79 \pm 0.01) \, {\rm GHz}$~\cite{Fixsen:2009ug}.
In the presence of a boost we can instead expand at second order in $\beta$ and first order in $\varepsilon$ getting:
\begin{equation}\label{eq:lin-temp}
    \frac{\delta I_\nu}{K} \,=\, \varepsilon \, \frac{\delta T(\boldsymbol{\hat{n}})}{T_0} +  \boldsymbol{{\beta}}\cdot \boldsymbol{\hat{n}}   +  (\boldsymbol{{\beta}}\cdot \boldsymbol{\hat{n}})^2 Q(\nu) -   \frac{1}{2} \beta ^2\,.
\end{equation}
This quantity was dubbed \emph{linearized temperature} in~\cite{Notari:2015kla}. Here we have discarded terms of order $\beta\, \varepsilon$ or higher, and we have defined the quantity
\begin{equation}
    Q(\nu) \,\equiv\, \frac{\nu}{2 \nu_0}  \coth \left(\frac{\nu }{2 \nu_0}\right) \,,
\end{equation}
as in~\cite{Kamionkowski:2002nd,Sunyaev:2013coa,Notari:2015kla}. So, in addition of a dipole correction we also have a frequency dependent quadrupole correction and a shift to the monopole. Splitting into solar and orbital motion we obtain
\begin{align}
\label{eqmain}
    \frac{\delta I_\nu}{K}  \,=\;\, &  \varepsilon \, \frac{\delta T(\boldsymbol{\hat{n}})}{T_0} + \boldsymbol{{\beta}_S}\cdot \boldsymbol{\hat{n}}  + Q(\nu) (\boldsymbol{{\beta}_S}\cdot \boldsymbol{\hat{n}}) ^2  + \boldsymbol{{\beta}_O}\cdot \boldsymbol{\hat{n}} + Q(\nu) (\boldsymbol{{\beta}_O}\cdot \boldsymbol{\hat{n}})^2   \nonumber \\
    & + 2 \, Q(\nu) (\boldsymbol{{\beta}_S}\cdot \boldsymbol{\hat{n}} ) ( \boldsymbol{{\beta}_O}\cdot \boldsymbol{\hat{n}}  )-  \beta_S \beta_O   -   \frac{1}{2} \beta_S^2  -   \frac{1}{2} \beta_O^2\,.
\end{align}

The leading time-dependent term, used for the calibration of WMAP~\cite{Hinshaw:2003fc,Hinshaw:2008kr} and Planck 2015~\cite{Adam:2015rua,Adam:2015vua} is the dipole $\boldsymbol{{\beta}_O}\cdot \boldsymbol{\hat{n}} $. The  Planck collaboration is  considering also all the above correction terms when performing the LFI 2013 and both
LFI and HFI 2015 calibration (see Eqs. (A.1) in~\cite{Aghanim:2013bta} and (5) in~\cite{Adam:2015vua}), but using $Q(\nu)=1$. However the appropriate correction factor $Q(\nu)$ is quite different from 1, mainly for HFI: it has a value of about $\{1.25,\,1.5,\,2.0,\,3.1\}$ respectively for the for the $\{100,\,143,\,217,\,353\}$ GHz channels.  The same does not apply to the two highest frequency channels (545 and 857 GHz) since they were calibrated on Uranus and Neptune using models for their atmospheric emission~\cite{Adam:2015vua}.

Note that since WMAP used lower frequency channels its $Q$ deviation from unity was smaller. The values are $\{1.01,\,1.03,\,1.04,\,1.09,\,1.22\}$ respectively for its $\{23,\,33,\,41,\,61,\,94\}$ GHz channels. In any case, WMAP had instrumental sensitivities much lower than Planck, so they did not even include the sub-dominant time-dependent terms in~\eqref{eqmain} in the first place.

Let us now comment on the possible effects of the relativistic frequency-dependent quadrupolar corrections with $Q(\nu)>1$.

\subsection{Doppler cross-terms}

The correction $2 \, Q(\nu) (\boldsymbol{{\beta}_S}\cdot \boldsymbol{\hat{n}} ) ( \boldsymbol{{\beta}_O}\cdot \boldsymbol{\hat{n}}  ) $  is only suppressed by a factor $2 Q(\nu) \beta_S\simeq (2 - 7)\times 10^{-3}$ compared to the leading order and, importantly, has the same time dependence (one year period). It should therefore be consistently included in the Planck calibration with the correct $Q(\nu)$ factor, important especially for the HFI maps.

It is difficult to assess  precisely the impact  of such a correction on the calibration factor, and subsequently on the CMB maps and on the individual multipoles released by Planck. At face value, compared to Planck 2015 which used $Q(\nu)=1$  it could represent a correction  on the gain factor that, although quite small for the LFI frequencies, it is up to $0.5\%$ on the 353 GHz channel. This would be of very similar size of the systematic errors as estimated by Planck: $\{0.35\%,\,0.26\%,\,0.2\%\}$ respectively for the three LFI channels at $\{30,\,44,\,70\}$ GHz, and  $\{0.09\%,\,0.07\%,\,0.16\%,\,0.78\%\}$ for the HFI channels at $ \{100,\,143,\,217,\,353\}$ GHz. Moreover such possible error could propagate on the maps with a similar or higher impact, since for instance the calibration change quoted in~\cite{Adam:2015rua} between the 2013 and 2015 release is about $0.8\%$ for LFI, $1\%$ for HFI at low frequencies and $2\%$ for the 353 GHz channel, and this has lead to a total shift of $2.3\%$ in the cosmological parameter $A_S \, e^{-2\tau}$ (see Table 1 of ~\cite{Planck:2015xua}).

However this would depend on the detailed calibration procedure. In fact if the instrument is calibrated within a short time period (of order a few hours) then the correction is basically time-independent and would constitute an unknown signal which would add to the primordial and foreground signals and which can be reconstructed iteratively by the calibration and map-making process itself,  following a  procedure which has been shown to converge and lead to an error of $0.1\%$ over the entire year in the gain factor in WMAP~\cite{Hinshaw:2003fc} after ${\cal O} (10)$ iterations, and an error of less than $0.5\%$ on a single ring \footnote{A ``ring'' in the Planck scanning strategy is the set of observations made during a period of fixed spin axis pointing of the instrument~\cite{Ade:2013sjv}.} after about 5 iterations in Planck~\cite{Ade:2013eta,Tristram:2011gq}. It is unclear therefore whether such correction could induce a ring-dependent bias in the gain, and how this would translate in an overall global gain factor.  In fact, note that if a map is built integrating over a time of order of a multiple of 1 year the effect could possibly average to something much smaller in a globally averaged gain factor.

Note that~\cite{Tristram:2011gq} claims that the bias on the Planck global gain factor due to ignoring the relativistic corrections to the orbital dipole  is actually much smaller and only of relative order $0.0006\%$, instead of ${\cal O}(0.1\%)$ (this has  been used in the 2013 release to justify the use of a non-relativistic approximation, see Appendix A.2 of ~\cite{Ade:2013eta}). Such suppression, shown on simulations, might be indeed due to averaging the effects over long time and many rings. However, from the inclusion of the relativistic terms in Planck 2015 calibration it seems this estimate may no longer be applicable to the current calibration procedure. It is thus not clear what could the effect be in Planck 2015 real data, so we stress that they should be properly included.\footnote{The new version of the Planck HFI Calibration paper~\cite{Adam:2015vua} also argues that the inclusion of frequency dependent coupling between solar and orbital dipoles is expected to have a much smaller impact than the above estimates. Nevertheless its exact impact remains to be quantified on calibration of the real data.}

The fact that such extra signal are time-dependent  could produce seasonal variations of the gain at the level of $0.1\%-0.5\%$ and also produce a spurious time-dependent quadrupole in the maps of order $\delta T/T\approx (2-6)\times 10^{-7}$ for HFI, which would also exhibit seasonal variations. It should therefore be possible to find seasonal variations in the gain and in the quadrupole of the real data in the 2015 release with amplitudes which grow with frequency and it should be possible to subtract them.

\subsection{Pure quadrupole terms}

The correction $Q(\nu) (\boldsymbol{{\beta}_O}\cdot \boldsymbol{\hat{n}})^2$ represents a different time-dependent effect, and is smaller than the previous one by an order of magnitude. In fact, the time period is only half a year since clearly one can rewrite the $\cos[t/\rm{year}]^2$ term as $1/2(1+\cos[2t/\rm{year}])$. If the calibration is averaged over exactly one year, the time-dependent signal would average down and we would be left with an inconsequential constant (it should not affect the estimate of the gain, because of the iterative procedure described above). If such an average is not carried out one should carefully take this into account with the correct $Q(\nu)$ factor.

The correction term  $Q(\nu) (\boldsymbol{{\beta}_S}\cdot \boldsymbol{\hat{n}})^2$ is actually the dominant relativistic signal, of order $(2-5)\times10^{-6}$, which adds to the quadrupole in the map-making procedure.  However, since it represents a time-independent additional signal it should not affect the estimate of the gain. This term has been separately studied  in~\cite{Kamionkowski:2002nd,Notari:2015kla} for a different reasons. In particular it modifies in a non-negligible way the statistical significance of claims of two CMB anomalies: the quadrupole-octupole alignment and the low-quadrupole value~\cite{Notari:2015kla}.

\subsection{Quadrupole leakage into the dipole}

A quadrupole can have a leakage into other multipoles in the presence of a mask, suppressed by the masked sky fraction $1-f_{\rm sky}$.  The leakage could induce  for instance a time-dependent dipole. For the Doppler cross-terms, it is of order $\delta T/T\approx (2- 7)\times 10^{-7} \times  \sqrt{1-f_{\rm sky}}$, which would constitute a relative correction of order ${\rm few} \times 10^{-5}$ on the measurements of the solar dipole.\footnote{In any case, it is not easy to forecast whether a small effect on several multipoles could build up and cause a non-negligible systematic effect to overall CMB parameters.}

For the quadrupole terms the leakage would be even bigger. It would result in a shift in the dipole of order of $(0.1\%-0.3\%)\times \sqrt{1-f_{sky}}$, which is a relative correction of order a few $0.01\%$, depending on the size of the mask and on the channel. Such a correction might imply a shift in the determination of the dipole of the {\it same} order of the present uncertainty in the Planck 2015 data, the nominal amplitude of which~\cite{Adam:2015rua} is $(3364.5 \pm 2.0)\mu K$ and direction $(l,b)=(264.00 \pm 0.03,48.24\pm 0.02)$. In more detail, it is separately estimated as $(3365.5 \pm 3.0)\mu K$ with $(l,b)=(264.01 \pm 0.05, 48.26\pm 0.02)$ for LFI and $(3364.5 \pm 1.0)\mu K$ with $(l,b)=(263.94 \pm 0.02, 48.21\pm 0.008)$ for HFI.

Another possible source of leakage is the aberration effect itself. Aberration couples neighboring multipoles, and therefore parts of the quadrupole can leak into the dipole. Not only the time-dependent  quadrupole will aberrate into a time-dependent dipole but the also the time-independent quadrupole part will, through $\boldsymbol{\beta_O}(t)$. These effects are however completely negligible as far as calibration goes, as they are both of order a few parts in $10^{-10}$ (considering that the quadrupole itself is low compared to the theoretical expectation).

\section{Discussion}

The Planck collaboration itself is now mentioning in the 2015 release~\cite{Adam:2015vua} that the relativistic terms must be included in the calibration, and so we stress that this should be done consistently using the $Q(\nu)$ factor. Such systematic error, even if by chance it averages to something small, can be easily corrected for and so it should be properly accounted for and precisely shown to be negligible or not in the final results. In fact, the need for relativistic terms seem to contradict previous claims that they are negligible~\cite{Tristram:2011gq}.

Regarding previous data, WMAP and the Planck 2013  release, the relativistic effects may also be a concern.  In fact for  WMAP  the calibration~\cite{Hinshaw:2003fc} is also based on the orbital dipole and presents a systematic error estimated to be of about $0.2\%$, which is of the same size of the relativistic corrections. Note that in this case however the frequency dependence is not worrisome, since all the channels have relatively low frequencies and so $Q(\nu)$ is very close to 1.
For the 2013 Planck release  the calibration was based on the solar dipole, using the measurement of WMAP. The first concern may be that a bias in the WMAP calibration may have propagated in Planck through the dipole. The second concern is that the quadrupolar correction $Q(\nu) (\boldsymbol{{\beta}_S}\cdot \boldsymbol{\hat{n}}) ^2$ mentioned above is a relative correction of order ${\cal O}(0.1\% - 0.3\% )$ to the non-relativistic calibration factor used in HFI 2013.

As pointed out in~\cite{Notari:2015kla} the factor $Q(\nu)$ must also be taken into account to correctly estimate the primordial quadrupole. It is important to note that if this is not taken into account, then also the quadrupole leakage into the dipole (and other multipoles) through the presence of a mask would not be properly corrected for. It would then give rise to a spurious dipole of order $\sqrt{1-f_{sky}}$ times a few parts in a thousand, and perhaps a bias in the gain factor of similar size. This is comparable to the current systematics in the measurement of the dipole itself~\cite{Adam:2015rua}.

The potentially most important impact of the corrections here discussed is in Planck \emph{polarization} data. For HFI 2015 that is still dominated by systematic residuals at large scales that are coming from temperature-to-polarization leakage. Such leakages include mismatch from gain uncertainty, which are relevant even at the $10^{-3}$ level~\cite{Adam:2015vua}, and as a consequence the Planck polarization maps at large scales cannot be  used yet for cosmology studies and were not included in the 2015 release. It is therefore crucial (and straightforward) to remove the frequency dependent relativistic corrections in the calibration and map-making procedure in order to be sure of its precise quantitative impact and to improve the control of systematics in the polarization data.

\acknowledgments
We thank Alessandro Gruppuso, Massimiliano Lattanzi, Paolo Natoli and Matthieu Tristram for useful discussions.

\bibliographystyle{JHEP2015}
\bibliography{calibration}

\end{document}